# Nonresonance adiabatic photon confinement in spherical mirror system. Experimental study

S.S. Popov[1], M.G. Atluhanov [1], A.V. Burdakov [1,2], M.U. Ushkova[3]

[1] Budker Institute of Nuclear Physics SB RAS 630090. Novosibirsk, Russia
[2] Novosibirsk State Technical University, 630090. Novosibirsk, Russia
[3] Novosibirsk State University, Novosibirsk, Russia
Email:S.S.Popov@inp.nsk.su

New nonresonance approach of photon accumulation in two spherical mirrors has been experimentally demonstrated. In this work, we have received a high accumulation coefficient and shown good effectiveness of this technique for creating photo neutralizer of negative ion beams. This efficiency in such concept is generally determined by reflectance mirrors and is practically not dependent of input emission quality and does not require high precision adjusting the optical elements.

**English introduction for Russian text**

One of the promising methods of neutralizing the negative ion beams is the application of photon target, in which electron reaction photodetachment was used [1].As known in photodetachment process additional reaction channels there are not. For example, in gas targets full ionization there is. This influence on the parameters of the neutral beam. In addition gas puffing impairs vacuum conditions. To implement the photon converter we must take into account that because of the small cross-section photodetachment the photons have to multiple cross the negative ion beam to produce a high degree of neutralization (90%). This can be achieved by using system of mirrors for accumulation and storing the radiation within it. Many different schemes have been proposed [1,2,3,4],but as a rule they are all based on different varieties of optical Fabry-Perot cavity. Systems of this type have a number of significant limitations. For flux multiplication about a thousand times needs not only mirrors with reflectivity of more than 0.999 and sufficient radiation resistance but and fulfillment of stringent requirements on radiation quality and adjusting optic elements. It needs to meet the phase-matching conditions for a large number of passes in resonator. In addition, a significant level of radiation power requires to take into account the temperature stabilization of the mirrors. Creating such a system based on modern technologies is possible, but requires a developmental work of resonator design. Current achievements and problems in the investigations of the resonator circuits can be found in [5,6].

Nonresonance accumulation of photons can be an alternative way [7,8]. This concept of the photon trap represents a reflective surfaces system which provides multiple rays return. The trap is similar to the mathematical billiards [8,9], which contains a sufficient volume of stability region in the phase space for the billiard trajectories. The energy density in this circuit increases in proportion of the lifetime of the rays. Integral photon lifetime in the trap is determined by the loss of photons due to reflection and due to time of escape from this system. This escape is inevitable, because it requires the input and output of the particle beam and the injection of a powerful flux. In difference from resonance cell this photon trap has not strict conditions for the phase relationships between a large number of rays inside and radiation input is feasible through a small hole or holes, but not through the high reflective surface. Conservation some of adiabatic invariants provides photon confinement. It turns out that effectiveness of the accumulation of photons is practically independent of the quality of the injected radiation. This circumstance allows us to consider as a source of radiation is sufficiently cheap industrial high-performance fiber laser.

In theory, the task of creating a non-resonant photon traps discussed in detail in [8] there were obtained on the requirements of the spatial configuration of the mirror system and the photon beam injection conditions.

In this work the experimental verification of the principle of non-resonant accumulation of radiant energy in a simple configuration of two spherical mirrors has been presented.

References

1. W. Chaibi, C. Blondel, L. Cabaret, C. Delsart, C. Drag, A. Simonin, Photoneutralization of negative ion beam for future fusion reactor. // E. Surrey, A. Simonin (Eds.), Negative Ions Beams and Sources: 1st International Symposium, AIP Conference Proceedings. — 2009. — Vol. 1097. — P. 385–394.
2. J.H. Fink, A.M. Frank. Photodetachment of electrons from negative ions in a 200 keV deuterium beam source. // Lawrence Livermore Natl. Lab. — 1975. — Report UCRL-16844.
3. J.H. Fink, Photodetachment technology, in: Production and Neutralization of Negative Ions and Beams: 3rd Int. Symposium, Brookhaven 1983, AIP, NewYork, 1984, pp. 547–560.
4. Vanek V., Hursman T., Copeland D., et al., Technology of a laser resonator for the photodetachment neutralizer. // Proc. 3rd Int. Symposium on Production and Neutralization of Negative Ions and Beams, Brookhaven. — 1983. — P.568-584.
5. M. Kovari, B. Crowley. Laser photodetachment neutraliser for negative ion beams// Fusion Engineering and Design. — 2010. — Vol. 85. — P. 745–751.
6. A.Simonin, L. Christin, H. de Esch, P. Garibaldi, C. Grand, F. Villecroze, C. Blondel, C. Delsart, C. Drag, M. Vandevraye et al. SIPHORE: Conceptual study of a high efficiency neutral beam injector based on photo-detachment for future fusion reactors // AIP Conf.Proc. 1390 (2011) 494-504.
7. С.С. Попов, М.Г. Атлуханов, А.В. Бурдаков и др. Нерезонансный фотонный нейтрализатор мощных пучков отрицательных ионов. Тезисы XLII международной конференции по физике плазмы и УТС. 9 – 13 февраля 2015 г. – 2015. С. – 395.
8. Popov S.S., Burdakov A.V, Ivanov A.A., Kotelnikov I.A. Preprint: http://arxiv.org/ftp/arxiv/papers/1504/1504.07511.pdf.
9. V. V. Kozlov, D. V. Treshchëv, Billiards. A genetic introduction to the dynamics of systems with impacts, Translations of Mathematical Monographs, 89, Amer. Math. Soc., Providence, RI, 1991, ISBN: 0-8218-4550-0 , viii+171 pp..

# Экспериментальное исследование нерезонансного накопления фотонов в системе сферических зеркал.

С.С. Попов[1], М.Г. Атлуханов[1], А.В. Бурдаков[1,2], М.Ю. Ушкова[3].

[1]*Институт ядерной физики им. Г.И. Будкера СО РАН 630090. Новосибирск, Россия*
[2]*Новосибирский государственный технический университет, 630090. Новосибирск, Россия*
[3]*Новосибирский государственный университет, Новосибриск, Россия*
*Email:S.S.Popov@inp.nsk.su*

*В работе экспериментально продемонстрирован новый, нерезонансный, подход к накоплению фотонов в системе двух сферических зеркал. В эксперименте получен высокий коэффициент накопления и показана достаточная эффективность данной методики для создания фотонейтрализатора пучков отрицательных ионов. Перспективность данной концепции определяется в основном отражательной способностью зеркал, практически не зависит от качества вводимого излучения и не требует сверхточной юстировки оптических элементов. Полученные в эксперименте данные согласуются с теоретическими расчетами.*

**Введение**

Во многих практических приложениях востребовано создание стационарных (квазистационарных) ансамблей фотонов высокой плотности. Традиционно для этого используются или предлагается использовать накопители излучения резонаторного типа, по сути эталонах Фабри-Перо [1]. Однако такой подход сопряжен с рядом трудностей: качество излучения накачки (ширина линии, эмиттанс), точность позиционирования, температурная стабилизация и др. В работе [2, 3] была рассмотрена возможность создания нерезонансных накопителей излучения, опирающаяся на адиабатически сохраняющиеся величины. Оказывается, такие инварианты позволяют ограничить занятую фотонами область, несмотря на наличие открытых участков зеркальной поверхности. При этом существенно ослабляются условия на качество излучения и точность юстировки зеркал. Реализация подобного подхода позволит существенно развить практические подходы в разных инженерных областях, требующих интенсивных фотонных потоков (необязательно когерентных): спектроскопия, оптическое разделение изотопов, фотохимии, фотонейтрализация пучков отрицательных ионов. Последнее направление является особенно перспективным для создания высокоэффективных генераторов мощных пучков нейтральных атомов в термоядерных приложениях таких как ИТЭР[], где КПД инжектора – один из ключевых параметров. Хотя в настоящее время имеется множество предложений по фотонным нейтрализаторам для мощных пучков, основанных на резонансном накоплении фотонов[4,5,6,7]. Все они далеки от практической реализации. Создание такой системы на базе современных технологий возможно, но требуется серьезная проработка элементов конструкции резонатора.

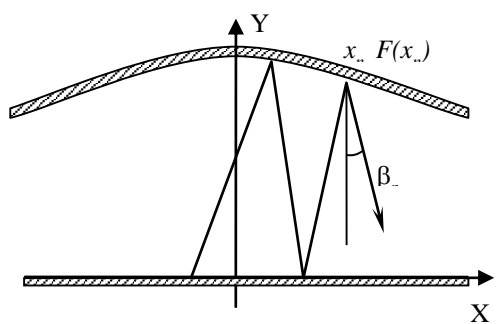

*Рис. 1. Схема адиабатической ловушки.*

Современные достижения и проблемы в изысканиях по резонаторным схемам можно найти в [7,8].

В [3] ловушка аналогична математическому бильярду Биркгофа [9] (с той разницей, что имеется открытые участки границы), который содержит достаточно объемную область устойчивых в фазовом пространстве бильярдных траекторий. Плотность энергии в этой схеме возрастет пропорционально времени жизни лучей. Интегральное время жизни в фотонной ловушке определяется, как и в резонансном накопителе фотонов, в основном, потерями фотонов на отражениях, а также временем их ухода за пределы системы, так как зеркала не могут образовать замкнутую поверхность, поскольку требуется место для ввода и вывода пучка частиц или другого объекта, а также ввода мощного потока излучения. Двумерная схема такой ловушки представлена на Рис. 1. Как видно из рисунка, при каждом отражении от верхнего зеркала фотон получает приращение горизонтального импульса в ту сторону, где расстояние $F$ до нижнего зеркала больше. В результате фотон будет отражаться от боковых торцов ловушки, где расстояние между зеркалами уменьшается, и возвращаться к положению «равновесия», если, угол $\beta_n$ между вертикалью и направлением движения достаточно мал. В [3] при рассмотрении уравнений движения

$$x_{n+1} - x_n = \left(F(x_{n+1}) + F(x_n)\right) tg\beta_n \qquad (1)$$

$$\beta_{n+1} - \beta_n = 2\frac{dF(x_{n+1})}{dx} \qquad (2)$$

было получено условие устойчивости квазипериодического движения фотона

$$F(0) < R, \qquad (3)$$

Которое наряду с адиабатическим инвариантом

$$F(x)\cos(\beta) = const \qquad (4)$$

позволяет рассчитать фотонные накопители заданных размеров.

Данная работа посвящена практической экспериментальной проверке принципа нерезонансного накопления лучистой энергии в простой конфигурации из двух сферических зеркал равной кривизны. Такая система эквивалентна приведенной на Рис. 1.

1. **Схема эксперимента**

Схема испытательного стенда для изучения метода представлена на Рис. 2. Ловушка состояла из двух оппозитных сферических зеркал

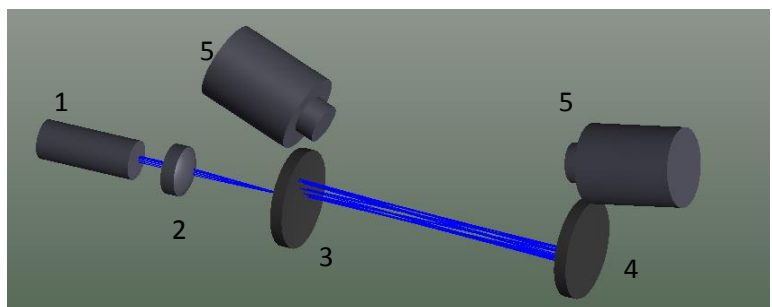

*Рис. 2. Схема эксперимента: 1 – источник излучения; 2 – линза; 3 – зеркало с входным отверстием; 4 – глухое зеркало; 5 – CCD-касмеры*

В качестве источника излучения (1) в работе применялся непрерывный YAG лазер (1.064 мкм) с диодной накачкой. Луч лазера вводился в систему двух сферических вогнутых зеркал с радиусом кривизны 250мм (3,4), через отверстие диаметром 300 мкм в центре одного из них. После чего фотоны испытывали некоторое количество отражений до выхода из системы или поглощения на зеркалах. Последние выполнены напылением около 40 диэлектрических слоев на кремниевую монокристаллическую подложку диаметром 2.5см. Пропускание контрольной пластины (свидетель) в диапазоне 1.03÷1.14 мкм не превышало 0.1%. Изготовление зеркал проведено в Институте лазерной физики СО РАН. Для анализа эффективности накопления лучистой энергии в ловушке регистрировалось паразитное рассеяние световых потоков на поверхности зеркал посредством CCD-камер (5) SDU 285

[10] с соответствующим набором светофильтров и объективов. Калибровка распределения чувствительности камер по полю зрения производилась с помощью равномерно освещенного матового экрана. Относительная калибровка чувствительности показала, что её отклонение носило монотонный характер и ограничивалось 2% в интервале регистрируемого поля зрения, а также согласовывалось с геометрическим удалением точек поверхности зеркала при движении по полю зрения. Также при подсчете коэффициента накопления, учитывался вклад фонового излучения.

В ходе эксперимента расстояние между зеркалами могло существенно меняться в пределах 50÷300 мм.

## 2. Измерение коэффициента накопления

Интегральная «яркость» картинки (см. Рис. 3), регистрируемоой камерами, очевидно пропорциональна общей мощности светового потока между зеркалами. В дальнейшем она сравнивалась с пятном рассеяния, полученным от первого отражения. Для этого отраженный луч на втором проходе поглощался с помощью светофильтра.

Для получения наилучшего коэффициента накопления фотонов, необходимо как можно дольше удерживать пучок в ловушке. Следовательно, задача сводится к поиску оптимальных условий для инжектируемого потока излучения при получении максимального соотношения суммарного потока излучения на зеркале (все отражения) к потоку первого отражения (см. Рис. 5, Рис. 6).

На Рис.3 представлена поверхность второго зеркала. Как видно, совокупность пятен рассеяния отражений пучка образует эллипс. Приближенно последовательность точек отражения на зеркале можно рассчитать в рамках работы [3]. Качественно такое поведение аналогично дискретному описанию колебаний материальной точки внутри сферической чаши в поле тяжести. Форма траектории определяется начальным моментом импульса фотона по отношению к оси симметрии системы. В частности, если момент импульса равен нулю, то проекция траекторий будет прямой линией. Если же вводить излучение с неравным нулю моментом импульса (или, что эквивалентно, прицельным параметром) относительно оси симметрии, то изображение в рассеянном свете представляет собой прецессирующий эллипс. При этом отражения будут происходить до тех пор, пока отраженный луч не попадет обратно во входное отверстие или полностью поглотится на многочисленных отражениях. На Рис. 2 и Рис. 3 приведены результаты измерений в условиях, когда эллипс быстро прерывается при попадании основной доли излучения в отверстие ввода.

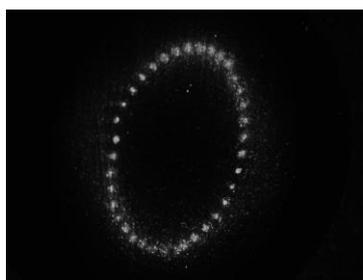

*Рис. 3. Снимок поверхности второго зеркала.*

С целью поиска оптимальных условий для наилучшего удержания излучения было выполнено моделирование эксперимента с помощью программы «Zemax». На Рис. 4. представлен расчетный след точек отражений на идеальных зеркалах. Их совокупность также образовывала прецессирующий эллипс. В результате такой прецессии след лучей на зеркале образует кольцо. При определенных углах инжекции входного излучения, можно было получить прямую линию или же фигуру близкую к окружности. В результате расчетов было определено оптимальное место ввода излучения, которое соответствует наименьшей плотности точек отражений. Из Рис.3 можно заключить, что оно должно находиться ближе к внешнему радиусу кольца, но все-таки не достигать его. Полученные результаты моделирования позволили существенно увеличить число отражений в ловушке. На Рис. 5. показана оцифрованная фотография с лучшим удержанием излучения.

Отметим что картина существенно отличается от ожидаемой на Рис. 4. Хорошо заметно 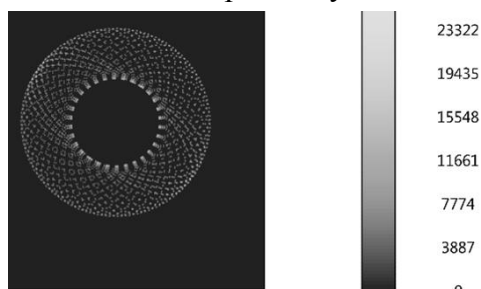 «заплывание» центра картинки, а также ограничение прецессии по азимутальному углу с весьма резкой границей. Причина, по-видимому, в несохранении момента импульса, вызванного небольшим отклонением зеркал от сферической формы. Наблюдаемое нарушение сохранение углового момента, может быть вызвано небольшим отклонением от сферичности формы зеркал. В этом случае, очевидно, отсутствует выделенная ось симметрии. На Рис.7 показан результат моделирования хода лучей в системе асферических зеркал с отклонением от сферической поверхности

*Рис. 4 Трассировка луча между идеальными сферическими зеркалами (порядка 8000 отр).*

порядка 10⁻⁵ м. Как видно, эллипс в ходе прецессии по часовой стрелке сжимается, что соответствует уменьшению момента импульса по абсолютной величине, и, как следствие, замедлению прецессии до ее последующего обращения в противоположную сторону. Результат моделирования качественно совпадает с наблюдаемым в эксперименте

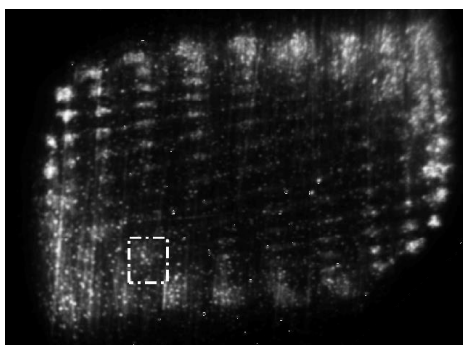

*Рис. 5. Пример профиля потока накопления лазерного излучения на глухом зеркале. Прямоугольный контур – область первого отражения*

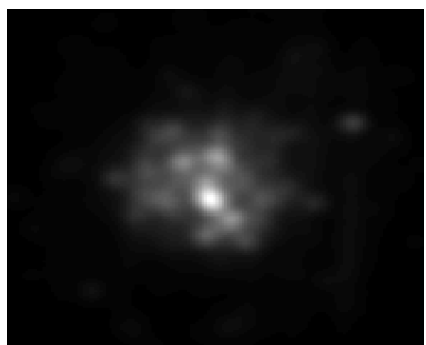

*Рис. 6. След первого прохода излучения*

распределением интенсивности. Из этого можно сделать вывод, что используемые зеркала также асферичны. Однако для конечной цели данного исследования это несущественно.

Для измерения эффективности накопления бралось удвоенное (для учета потока туда-обратно) отношение интеграла профиля потока накопленного излучения на Рис. 5. к интегралу от первого прохода на Рис. 6. Наилучшее значение эффективности накопления (отношение накопленной мощности излучения ко входной) составило величину 315±10. Это значение соответствует коэффициенту отражения R≈0.9968. Вероятно, существенная часть излучения все же выходила в отверстие ввода. Полностью этого избежать нельзя вследствие теоремы Пуанкаре о возвращении [11]. Из-за этого эффективность накопления оказалась меньше чем величина $10^3$, получаемая из заявленного производителем коэффециента отражения R≥0.999. Опыты были поставлены в пределах зоны устойчивости (3).

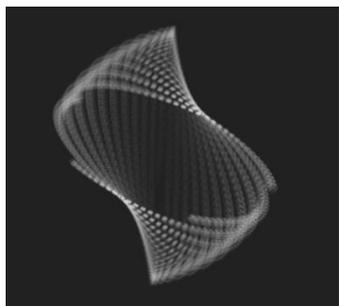

*Рис.7. Результат моделирования точек рассеяния в системе асферических зеркал.*

**Заключение**

Проведенные эксперименты показали перспективность нерезонансного накопления излучения для задач, требующих высокой плотности лучистой энергии в значительных по сравнению с длиной волны пространственных объемах. Проверенный подход очевидно нечувствителен к качеству излучения накачки, точности позиционирования и стабилизации оптических элементов. Известные авторам схемы реализованных резонаторных накопителей с выдающимися параметрами [12,13] имеют маленькие размеры и малую плотность энергии, что затрудняет их практическое применение. Предложенный же в [2,3] и апробированный здесь подход имеет широкие практические перспективы – спектроскопия, лазерное разделение изотопов, фотонейтрализация пучков отрицательных ионов.

**Благодарности**

1 G. Rempe, R. J. Thompson, H. J. Kimble, and R. Lalezari, Optics Letters, Vol. 17, Issue 5, pp. 363-365 (1992)
2 С.С. Попов, М.Г. Атлуханов, А.В. Бурдаков и др. Нерезонансный фотонный нейтрализатор мощных пучков отрицательных ионов. Тезисы XLII международной конференции по физике плазмы и УТС. 9 – 13 февраля 2015 г. – 2015. С. – 395.
3 Попов С.С., Бурдаков А.В., Иванов А.А., Котельников И.А. Препринт: http://arxiv.org/ftp/arxiv/papers/1504/1504.07511.pdf.
4 J.H. Fink, A.M. Frank. Photodetachment of electrons from negative ions in a 200 keV deuterium beam source. // Lawrence Livermore Natl. Lab. — 1975. — Report UCRL-16844.
5 J.H. Fink, Photodetachment technology, in: Production and Neutralization of Negative Ions and Beams: 3rd Int. Symposium, Brookhaven 1983, AIP, NewYork, 1984, pp. 547–560.
6 Vanek V., Hursman T., Copeland D., et al., Technology of a laser resonator for the photodetachment neutralizer. // Proc. 3rd Int. Symposium on Production and Neutralization of Negative Ions and Beams, Brookhaven. — 1983. — P.568-584.
7 M. Kovari, B. Crowley. Laser photodetachment neutraliser for negative ion beams// Fusion Engineering and Design. — 2010. — Vol. 85. — P. 745–751.
8 A. Simonin, L. Christin, H. de Esch, P. Garibaldi, C. Grand, F. Villecroze, C. Blondel, C. Delsart, C. Drag, M. Vandevraye et al. SIPHORE: Conceptual study of a high efficiency neutral beam injector based on photo-detachment for future fusion reactors // AIP Conf.Proc. 1390 (2011) 494-504.
9 В.В. Козлов, Д.В. Трещев. Биллиарды. Генетическое введение в динамику систем с ударами. – М: Изд-во МГУ, 1991. – 168 с.
10 http://www.sptt.ru/sptt/catalog.php?mod=sdu285.
11 Арнольд В. И. Математические методы классической механики. Изд. 5-е стереотипное. — М.: Едиториал УРСС. — 2003. — С. 62. — ISBN 5-354-00341-5.
12 G. Rempe, R. J. Thompson, H. J. Kimble, and R. Lalezari Optics Letters, Vol. 17, Issue 5, pp. 363-365 (1992)
13 Christina J. Hood, H. J. Kimble, and Jun Y. Characterization of high-finesse mirrors: Loss, phase shifts, and mode structure in an optical cavity.// Phys. Rev. A 64, 033804 (2001).